\documentclass[10pt,conference]{IEEEtran}

\IEEEoverridecommandlockouts
\usepackage{amsmath,amssymb,amsfonts}
\usepackage{algorithmic}
\usepackage{cite}
\usepackage{graphicx}
\usepackage{textcomp}
\usepackage{xcolor}

\usepackage[bottom]{footmisc}
\usepackage{hyperref}
\usepackage{footnotebackref}
\usepackage{cleveref}
\usepackage{booktabs}
\usepackage{subcaption}
\usepackage{listings}
\usepackage{courier}
\usepackage{array}
\usepackage{verbatim}
\usepackage{caption}
\usepackage{cancel}
\usepackage{multirow}
\usepackage{stmaryrd}
\usepackage{enumitem}
\usepackage{framed}
\usepackage{xargs}
\usepackage{tabularx}

\renewenvironmentx{leftbar}[2][1=1pt, 2=5pt]% 
{\def\FrameCommand{\vrule width #1 \hspace{#2}}\MakeFramed {\advance\hsize-\width \FrameRestore}}% 
{\endMakeFramed}

{\begin{leftbar}\begin{quotation}}%
{\end{quotation}\end{leftbar}}

\def\BibTeX{{\rm B\kern-.05em{\sc i\kern-.025em b}\kern-.08em
    T\kern-.1667em\lower.7ex\hbox{E}\kern-.125emX}}

\newcommand{\hackctx}[0]{\textit{HackIde}}
\newcommand{\hackselec}[0]{\textit{HackAutocompletion}}
\newcommand{\hackall}[0]{\textit{HackAll}}
\newcommand{\pythonctx}[0]{\textit{PythonIde}}
\newcommand{\pythonselec}[0]{\textit{PythonAutocompletion}}
\newcommand{\pythonall}[0]{\textit{PythonAll}}
\newcommand{\multictx}[0]{\textit{MultiIde}}
\newcommand{\multiselec}[0]{\textit{MultiAutocompletion}}
\newcommand{\multiall}[0]{\textit{MultiAll}}
\newcommand{\diff}[0]{\textit{HackCommit}}

\newcommand{\trans}[0]{\textit{GPT-2}}
\newcommand{\snippetbart}[0]{\textit{BART}}

\begin{document}

\title{Improving Code Autocompletion with Transfer Learning}

\author{
    \IEEEauthorblockN{Wen Zhou}
    \IEEEauthorblockA{
        \textit{Facebook Inc.}\\
        Menlo Park, U.S.A. \\
        zhouwen@fb.com
    }
    \and
    \IEEEauthorblockN{Seohyun Kim}
    \IEEEauthorblockA{
        \textit{Facebook Inc.}\\
        Menlo Park, U.S.A. \\
        skim131@fb.com
    }
    \and
    \IEEEauthorblockN{Vijayaraghavan Murali}
    \IEEEauthorblockA{
        \textit{Facebook Inc.}\\
        Menlo Park, U.S.A. \\
        vijaymurali@fb.com
    }
    \and
    \IEEEauthorblockN{Gareth Ari Aye}
    \IEEEauthorblockA{
        \textit{Facebook Inc.}\\
        Menlo Park, U.S.A. \\
        gaa@fb.com
    }
}

\maketitle
% \IEEEpeerreviewmaketitle

\begin{abstract}
Software language models have achieved promising results predicting code completion usages, and several industry studies have described successful IDE integrations. Recently, accuracy in autocompletion prediction improved 12.8\%\cite{aye2020learning} from training on a real-world dataset collected from programmers' IDE activity. But what if limited examples of IDE autocompletion in the target programming language are available for model training? In this paper, we investigate the efficacy of pretraining autocompletion models on non-IDE, non-autocompletion, and different-language example code sequences. We find that these unsupervised pretrainings improve model accuracy by over 50\% on very small fine-tuning datasets and over 10\% on 50k labeled examples. We confirm the real-world impact of these pretrainings in an online setting through A/B testing with thousands of IDE autocompletion users, finding that pretraining is responsible for increases of up to 6.63\% autocompletion usage.
\end{abstract}

\begin{IEEEkeywords}
Machine learning, neural networks, software language models, naturalness, code completion, integrated development environments, software tools
\end{IEEEkeywords}

\section{Introduction}
\begin{figure}
\includegraphics[width=0.95\linewidth]{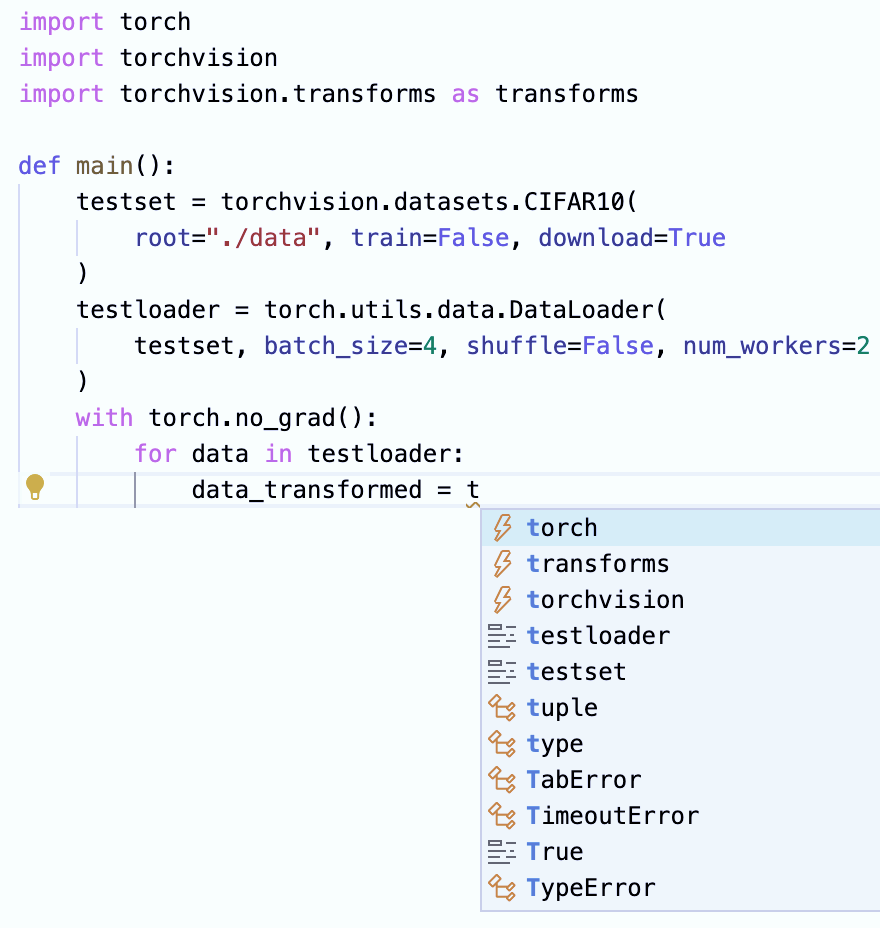}
\caption{Example of autocomplete in an IDE. The first 3 suggestions with thunderbolt icons are provided by our deep learning model.}
\label{fig:ide-example}
\end{figure}

Autocompletion is the most frequently used IDE feature\cite{Murphy:2006:JSD:1159169.1159396}. Significant attention has been given to improving suggestion
prediction through machine learning\cite{Bruch:2009:LEI:1595696.1595728,Hindle:2012:NS:2337223.2337322,aye2020sequence,kim2020code} by feeding code
to models as a sequence of tokens or even AST nodes \cite{Li_2018}. Figure \ref{fig:ide-example} shows an example of autocomplete powered by deep learning in an IDE. Several recent studies \cite{aye2020learning,hellendoorn-codefails} have demonstrated the strengths of real-world and weaknesses of synthetic datasets in training and evaluating autocompletion models. Concept drift between real-world and synthetic examples can be ameliorated by only showing models real-world autocompletion selections. But what if autocompletion examples from IDE usage in the target language are hard to come by?

Even with a large user population, the tool's vendor may not be able to log autocompletion events to build a training dataset. But perhaps the vendor is able to collect random, non-autocompletion code sequences during code authoring. These would suffice to train a task-agnostic language model (LM) for autocompletion prediction, but under the assumption that tokens used in  autocompletion follow the same distribution as tokens in arbitrary code sequences. One flaw in this assumption is that modern IDE autocompletion tools restrict the token types which can be suggested, commonly disallowing punctuation, literals, and variable declarations. Furthermore, several recent studies\cite{aye2020learning,hellendoorn-codefails} have shown a myriad of differences between random tokens sampled from source code and those used in autocompletion. Can knowledge from these unlabeled code sequences transfer to the autocompletion task?

\begin{figure}
\centering
\includegraphics[width=0.95\linewidth]{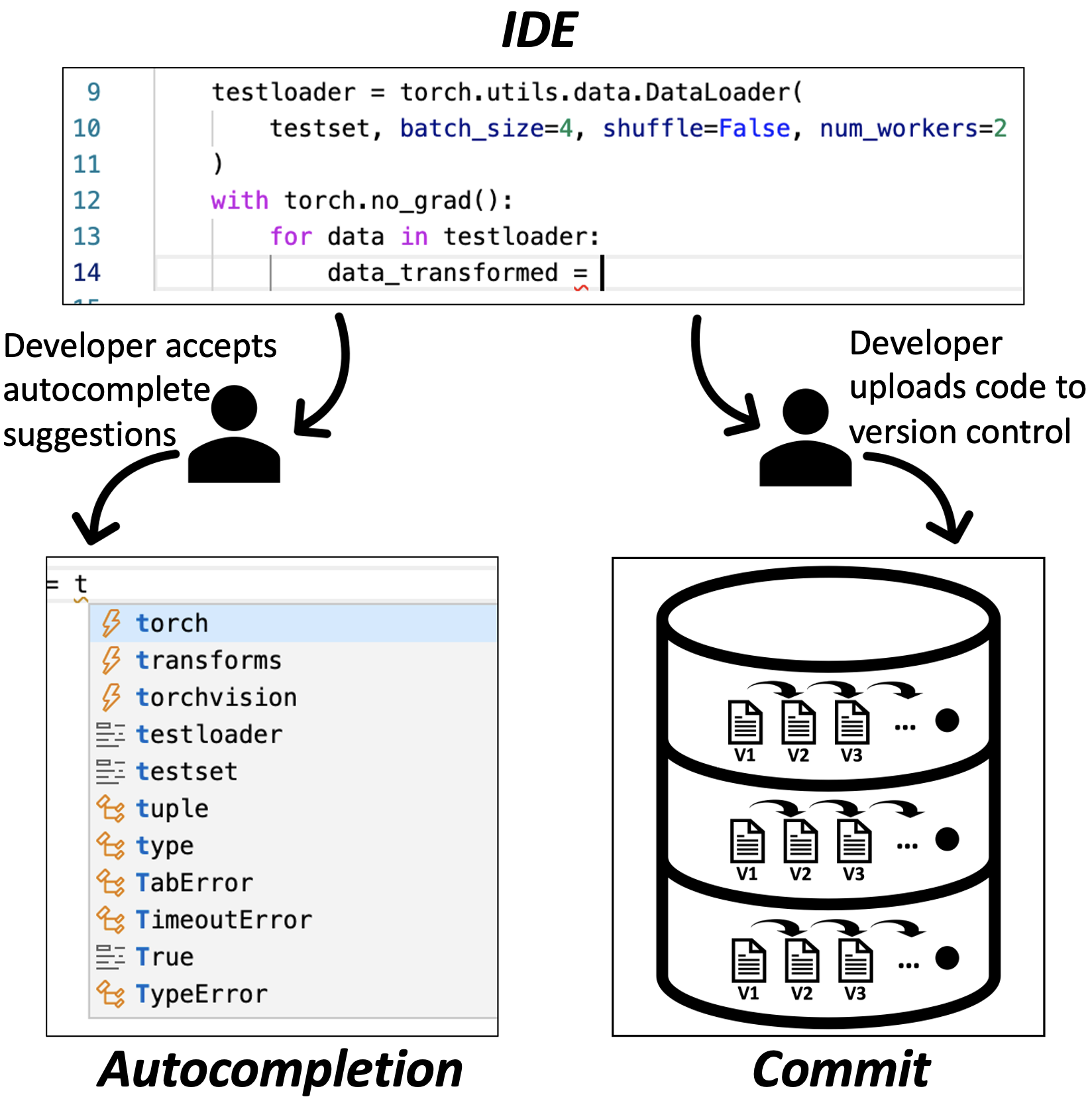}
\caption{Different code authoring stages that could be used for autocompletion training data. IDE dataset consists of snapshots of source code files collected during code authoring. Once a developer submits the code to code review, it becomes part of Facebook's version control data. From the IDE, a developer can also choose an autocompletion suggestion, which would be logged (along with the surrounding context code).}
\label{fig:code-cycle}
\end{figure}

Consider the various stages of code that could be leveraged for autocompletion model training. Figure \ref{fig:code-cycle} shows the different code datasets. First, we have code as it appears in the IDE$\textendash$snapshots of code taken from real-time code authoring. Branching from there, two events that create additional datasets are autocompletion and commit upload. The former is when a developer accepts an autocompletion suggestion in the IDE. The selection, list of suggestions, and surrounding context are logged. The latter is when a commit is uploaded to version control. The commit contains a snapshot of each of the impacted files. There is an intuitive relationship between code commits and developers' IDE activity since a commit is the first code artifact that a developer will produce after code authoring. Can knowledge from these commits transfer to modeling code authoring behaviors?

\begin{figure}
\centering
\includegraphics[width=\linewidth]{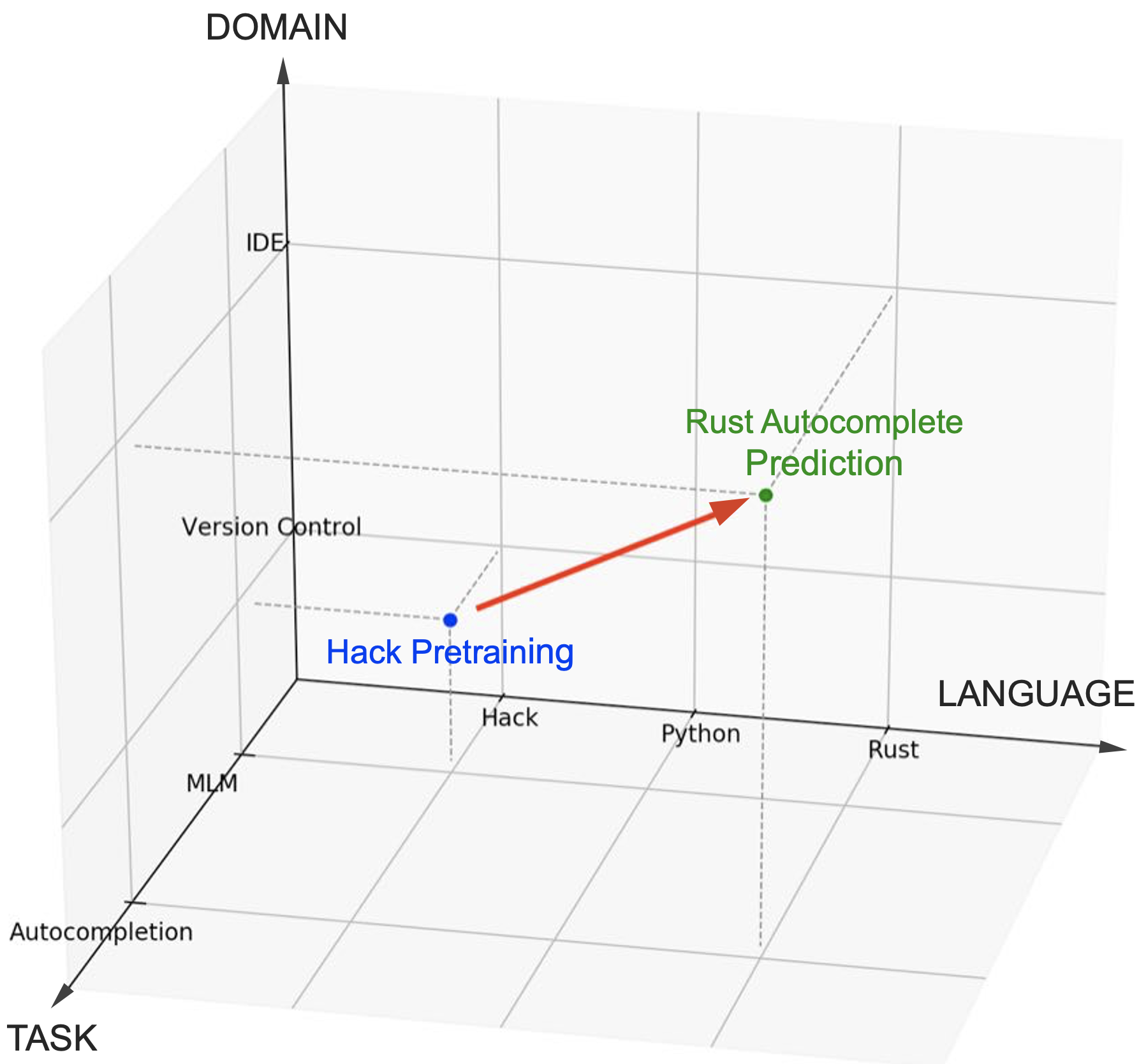}
\caption{3D illustration of the transfer learning space. We explored transfer learning across tasks (RQ1), domains (RQ2), and languages (RQ3) in this paper.}
\label{fig:transfer-space}
\end{figure}

Another potential cause for insufficient labeled data is a non-uniform programming language usage distribution. While there are undeniably shared concepts and constructs across programming languages, there are often major differences in syntax, programming style, and language constructs. Additionally, language-specific differences in IDE tools may impact programmers' autocompletion behaviors. At Facebook, we have a large developer activity dataset in the Hack programming language containing millions of real-world autocompletion examples. But there is also a long tail of less frequently used languages such as Rust. These languages lack enough labeled data for robust model training, but they are still used by a sufficient developer population to warrant autocompletion support. Can knowledge from the more popular languages transfer to others where labeled data is insufficient?

In this paper, we explore these questions regarding transfer learning for autocompletion empirically. Advances in Transformer-based neural models~\cite{radford2019language,devlin2018bert,lewis-etal-2020-bart,feng2020codebert,ahmad2020summarization} have popularized transfer learning in the deep learning community. Often transfer learning consists of ``pre-training'' such a model on large corpora of unlabeled data in an unsupervised manner and then ``fine-tuning'' on a smaller corpus of labeled data. The latter fine-tuning corpus is typically drawn from the same distribution as the test dataset. In our study, the evaluation task is to predict IDE autocompletion selections made by programmers in a given language, and our fine-tuning datasets consist of real-world autocompletion events collected from IDE usage at Facebook. As visualized in Figure \ref{fig:transfer-space}, we explore the effect of transfer learning by pretraining models on non-IDE, non-autocompletion, and different programming language code sequences. Specifically, we answer the following research questions:

\begin{leftbar}
\noindent \textit{RQ1: How do autocompletion models benefit from combining unsupervised pretraining with task-specific fine-tuning? How does their performance improve across offline and online evaluation?}
\end{leftbar}

Our experimental results show that pretraining on code sequences collected during code authoring and fine-tuning on tokens selected through autocompletion produces models which outperform training on only one of these two datasets. We show further that such a model drives greater online tool usage.

\begin{leftbar}
\noindent\textit{RQ2: What is the impact of pretraining on a large source code dataset obtained from outside of code authoring? Can these pretrained software language models be fine-tuned on IDE autocompletion to achieve better accuracy with fewer real-world examples?}
\end{leftbar}

We show that starting from a model pretrained on files appearing in version control commits drastically decreases the number of real-world autocompletion examples required to achieve high accuracy. Additionally, our experiments suggest that there is diminishing marginal benefit to pretraining as the number of real-world examples grows.

\begin{leftbar}
\noindent\textit{RQ3: Consider the case where a large training corpus is available in one language but not another. Can pretraining a multilingual model on the language with more training data benefit the language with less data?}
\end{leftbar}

To answer this question, we pretrain a multilingual model on examples from one language before fine-tuning on examples from another. Our results show that many fewer target-language examples are required to achieve high accuracy after pretraining on different-language examples. We again observe diminishing marginal benefit to pretraining on different-language examples as the number of available target-language examples grows.\\

\noindent\textbf{Contributions}
\begin{enumerate}%[leftmargin=1.5em]
    \item We pretrain two transformer software language models GPT-2\cite{radford2019language} and BART\cite{lewis-etal-2020-bart} on source code files obtained from version control commits and show how their performance on autocompletion prediction improves through fine-tuning on real-world IDE code sequences.
    
    \item The GPT-2 model is trained on two real-world datasets: code sequences logged during IDE authoring and autocompletion selections. A third variant is pretrained on the former and fine-tuned on the latter corpus to demonstrate how the combination of pretraining and task-specific fine-tuning lead to a superior model, outperforming the base model by \textbf{3.29\%}.
    
    \item We show that pretraining on a different programming language boosts accuracy by \textbf{13.1\%} when comparing a model pretrained on Hack examples and fine-tuned on 10k Python examples versus only training on Python examples.
    
    \item We prove that improvements across these three transfer learning dimensions\textemdash task, domain, and language \textemdash translate into increased autocompletion tool usage by \textbf{3.86\%}, \textbf{6.63\%}, and \textbf{3.62\%}, respectively, comparing these models through online A/B tests.
\end{enumerate}

\vspace{2mm}

\noindent\textbf{Outline}

The rest of this paper is organized to first introduce our experimental setup in Section II. In this section we describe the corpora, language models, and evaluation methodologies employed in our study. Section III reports experimental results, supporting the answers to our research questions. Section IV, Section V, and Section VI discuss threats to validity, related work, and future work, respectively. Finally, Section VII concludes with a summation and key insights for autocomplete tool designers.

\section{Experimental Setup}
\label{sec:experiments}

\subsection{Datasets}
%\todo{we might want to separate the actual datasets from where we take losses to make it %cleaner}
This study's datasets (summarized in Table \ref{tab:dataset}) are sourced from real-world developer activity at Facebook. We focus on two languages, Hack and Python, with significant developer populations at Facebook. Going forward, we will refer to each dataset by \emph{[language][dataset]} (e.g. \diff{}). Figure \ref{fig:code-cycle} shows a diagram of the datasets as part of the various stages of the code cycle. 

\begin{enumerate}
    \item \emph{Autocompletion}: Autocompletion events logged whenever a programmer accepts a completion suggestion. The accepted suggestion, other unused suggestions (from static analysis), and the surrounding program context are logged.
    \item \emph{IDE}: Snapshots of source code files collected during code authoring. In addition to capturing the file contents, we collect the cursor position so that specific code fragments undergoing modification can be identified. This dataset shares the code authoring domain with \emph{Autocompletion}, but while \emph{Autocompletion} is task-specific, \emph{IDE} is importantly task-agnostic.
    \item \emph{Commit}: The set of files created or modified in a version control commit and uploaded to Facebook's code review service. This version control dataset is included in our investigation to explore transfer learning across domains in light of the concept drift reported in \cite{aye2020learning} and \cite{hellendoorn-codefails}. In a typical programming workflow, version control commits represent the software artifacts with the closest relationship to IDE code authoring.
    \item \emph{All}: A union of \emph{Autocompletion} and \emph{IDE} constructed to explore the effect of mixing labeled data into pretraining.
\end{enumerate}

It's important to note that the datasets we explore contain minimal overlap since they were collected from different time ranges. Major differences between these three datasets are catalogued in ~\cite{aye2020learning}. Furthermore, training and testing datasets were split by a random cutoff period, ensuring no data leaks between the datasets.

Additionally, a valid concern is raised in \cite{allamanis2019adverse} regarding the potential of duplicate examples in source code models split between training and evaluation datasets due to the prevalence of copy-paste in software development. In a nutshell, model accuracy may be overestimated if many held-out test examples appear in the model's training data owing to code clones. Luckily, this issue does not apply in our evaluation since the test dataset consists of real-world autocompletion events. Unlike fragments of source code, programmers will never copy autocompletion events. Furthermore, all of the improvements we attribute to transfer learning are confirmed in online A/B tests, so increased model performance cannot be caused by flaws in our held-out test dataset.

We train a variety of monolingual models on examples from only one of Hack and Python as well as several multilingual models on a union of examples from both languages. When training on a union of Hack and Python examples, we construct the model vocabulary \emph{V} = $V_{hack} \cup V_{python}$. Code sequences fed to our multilingual model are modified by prepending a control code\cite{keskar2019ctrl} to indicate the source language.

\begin{table}[]
\caption{Various datasets used in this paper.}
\def\arraystretch{1.20}
\begin{tabular}{|l|l|l|l|l|}
\hline
             & \multicolumn{2}{c|}{Python} & \multicolumn{2}{c|}{Hack} \\ \hline
  & \# tokens  & \# samples  & \# tokens & \# samples \\ \hline
\emph{Autocompletion}&   543,401,684   & 3,201,299      & 642,886,605  & 3,792,485     \\ \hline
\emph{IDE}        &   540,200,385    & 3,201,299      &   639,094,120    & 3,792,485     \\ \hline
\emph{Commit}         &    -        &      -        & 629,192,335          & 3,452,434     \\ \hline
\end{tabular}
\label{tab:dataset}
\end{table}

\subsection{Tokenization}

One difficulty in modeling source code as compared to natural language is that code introduces
new vocabulary at a far higher rate \cite{DBLP:journals/corr/abs-1903-05734}. Recognizing and predicting rare and novel tokens from an open vocabulary poses a challenge when our models are trained to recognize a fixed set of terms. A strategy explored in \cite{DBLP:journals/corr/abs-1903-05734} is to model code as a sequence of partial tokens. This scheme allows an open vocabulary of code tokens to be represented using a fixed-size vocabulary. Another idea from \cite{Li_2018} is copy mechanism where out-of-vocabulary (OOV) tokens are encoded so that model can recognize matches and predict these terms by reference. In this study we tokenize code fragments in two different ways:

\paragraph{Byte-pair encoding (BPE)} This scheme used in our \snippetbart{} model tokenizes source code tokens as a sequence of partial tokens. Common character sequences are encoded with a single subword whereas rare character sequences are encoded with longer sequences of shorter subwords. BPE has been applied successfully across natural language \cite{DBLP:journals/corr/SennrichHB15} and programming language \cite{DBLP:journals/corr/abs-1903-05734} modeling.
\paragraph{Bigram encoding + copy mechanism} This scheme used in our \trans{} model tokenizes snake\_case and camelCase identifier tokens as exactly two subtokens (bigrams). We selected this encoding strategy for online experiments in autocompletion ranking since it yields a short, fixed-height partial token tree to search during inference. In our online deployment, we worked with a 100ms latency budget for model predictions. Searching a height-2 tree is fast enough to meet low latency user experience requirements in autocompletion.

Concretely, consider a vocabulary \emph{V} of partial tokens and an example token \emph{t} = \lstinline{``fooBarBazQuux"}. First \emph{t} is broken into a list of partial tokens \lstinline{[``foo", ``Bar", ``Baz", ``Quux"]}. Then we choose a split point that cuts the list into two equal length sublists and join the elements of each list to form a bigram ($b_1, b_2$) = \lstinline{(``fooBar", ``BazQuux")}. Finally the $i^{th}$ unique, OOV bigram $b_i$ $\notin$ \emph{V} is replaced with a synthetic token \lstinline{<var-i>} as in \cite{DBLP:journals/corr/abs-2010-12663}. The special case of \emph{t} $\in$ \emph{V} receives a length-2 encoding of \lstinline{[t, </t>]} where \lstinline{</t>} is a synthetic end-of-token identifier.

%An example of a code snippet with the different tokenization methods can be seen in Table %~\ref{tab:encoding-example}.

%\begin{table}[]
%\begin{tabular}{|ll|}
%\hline
%Code                                                          & \begin{tabular}[c]{@{}l@{}}if %a.b():\\     run\_something()\end{tabular} \\ \hline
%\begin{tabular}[c]{@{}l@{}}Bigram \\ encoding\end{tabular}    &                                                                          \\ \hline
%\begin{tabular}[c]{@{}l@{}}Byte-pair \\ encoding\end{tabular} &                              %                                            \\ \hline
%\end{tabular}
%\label{tab:encoding-example}
%\caption{Different types of token encoding for source code}
%\end{table}

\subsection{Models}
In this paper, we evaluate the effects of transfer learning using two models, both incorporating Transformer architectures. Since the main focus of this paper is to examine the impact of transfer learning, we limit our focus to these two models and do not compare their performance to other state-of-the-art models. Our experiments leverage: 
\paragraph{\trans{}} a decoder transformer model ~\cite{radford2019language}, which has achieved state-of-the-art performance in code prediction \cite{kim2020code} due to the transformer's ability to observe all sequence elements simultaneously in its self-attention blocks.
\paragraph{\snippetbart{}} a bidirectional model that utilizes an encoder to encode the context surrounding the code completion point, as well as a GPT-2-like decoder for auto-regressive generation \cite{lewis-etal-2020-bart}. It is trained on a denoising objective. \snippetbart{} demonstrates state-of-the-art performance across a variety of source code modeling tasks in  \cite{ahmad2020summarization}. 
    % \todo{fill in and maybe include a diagram of the model?}

%\wen{We used bigram encoding + copy mechanism for \trans{} and used BPE encoding for %\snippetbart{} in this study. Do we need to explain why here?}

\subsection{Training}
Each of the software language models were trained in two phases. The first phase is pretraining, in which models are shown a large number of source code examples drawn from a dataset with concept drift from the evaluation dataset. The second phase is fine-tuning where models are shown source code examples drawn from the same distribution as the evaluation dataset. Some models are fine-tuned twice (e.g. row 3 of Table~\ref{tab:ab-results}) to achieve knowledge transfer across two different axes). All models were trained for up to twenty epochs (with early termination at convergence) using Nvidia Tesla V100 GPUs. The learning rates for pretraining and fine-tuning were set to $5^{-4}$ and $5^{-6}$ respectively\footnote{We reduced the fine-tuning learning rate after observing models abandoning solutions discovered during pretraining and overfitting the smaller fine-tuning datasets.}.

%First there is a base training step which can be used as a learning phase for a general task, %and then a fine-tuning step, which is more specific to the evaluation task. The fine-tuning %step occurred only on the autocompletion selection dataset (i.e. \hackselec{} or %\pythonselec{}) as it represented the evaluation task the best.
%All the models were trained using Nvidia Tesla V100 GPUs with max number of epochs at 20 and %early stopping of 2 epochs. The learning rate for the base training and fine-tuning was 5e-4 %and 5e-6 respectively\footnote{The learning rate for fine-tuning is smaller to not overfit in %the first couple epochs}. 

\subsection{Evaluation}
We measure the performance of our study's models through both offline and online evaluation.

\paragraph{Offline evaluation}
For offline evaluation, we use 10\% of \hackselec{} and \pythonselec{} as held-out test datasets. The evaluation task is for our autocompletion models to predict users' real, historical IDE autocompletion selections given the surrounding program context and suggestion list (from static analysis). \hackselec{} examples have an average of 99.5 candidate suggestions to choose from and \pythonselec{} examples have an average of 26.3. The candidate suggestions list allows us to frame our evaluation as a ranking problem. For each offline measurement, we report top-1 and top-3 accuracy as well as mean reciprocal rank at \emph{k} = 3 (MRR@3). MRR is defined as:
\begin{equation}
\text{MRR} = \frac{1}{n}\sum_{i=1}^{n}\frac{1}{rank_{i}}
\end{equation}
where $n$ is the size of test dataset and $rank_i$ is the rank of the correct token predicted by the model as the $i^{th}$ candidate. In our evaluation, we only consider the top \emph{k} = 3 results (otherwise the score will be zero).

\paragraph{Online evaluation}
The ultimate goal of our research is to improve the developer IDE experience. While offline evaluation is faster and lower-cost, it's imperative to test improvements with real users. In this study, we ran several live A/B experiments with thousands of Hack developers at Facebook. In each experiment, developers are randomly assigned to an experiment or control group, and we measure daily completions per user (DCPU). DCPU refers to the raw number of autocompletion suggestions a developer accepts on a given day. Using this metric, A/B test observations are taken as the number of times a given developer in one of these two groups uses autocompletion on a given day. We conduct each experiment until it reaches statistical significance of at least 95\%\footnote{Each experiment takes approximately two weeks to reach statistical significance.}.
\section{Results}
\label{sec:results}

\begin{table*}[]
\caption{Production A/B test results. The evaluation metrics is DCPU - daily completions accepted per user. Our threshold for p-value is 0.05.}
\centering
\def\arraystretch{1.20}
\small
\begin{tabular}{c|c|c|c|c|c}
\toprule
  & Pretraining & Fine-tuning   & \# unique developers & improvement & p-value \\
\midrule
%%% ---------------------------------------- hack ---------------------------------------- %%%
\multirow{3}{*}{1} & \hackctx{} & -   & 3912 & - & - \\
\cline{2-6}
 & \hackctx{} & \hackselec{}   & 3933 & 0.0386 & 0.0238 \\
\midrule
\multirow{3}{*}{2} & - & \hackselec{}   & 3002 & - & - \\
\cline{2-6}
 & \hackctx{} & \hackselec{}  & 3022 & 0.0663 & 0.0172 \\
\midrule
%%% ---------------------------------------- python ---------------------------------------- %%%
\multirow{3}{*}{3} & - & 100k (\hackall{} $\rightarrow$ \hackselec{})   & 3704 & - & - \\
\cline{2-6}
 & \diff{} & 100k (\hackall{} $\rightarrow$ \hackselec{})  & 3697 & 0.0362 & 0.0494 \\
\bottomrule
\end{tabular}
\label{tab:ab-results}
\end{table*}

\begin{figure}
\includegraphics[width=0.95\linewidth]{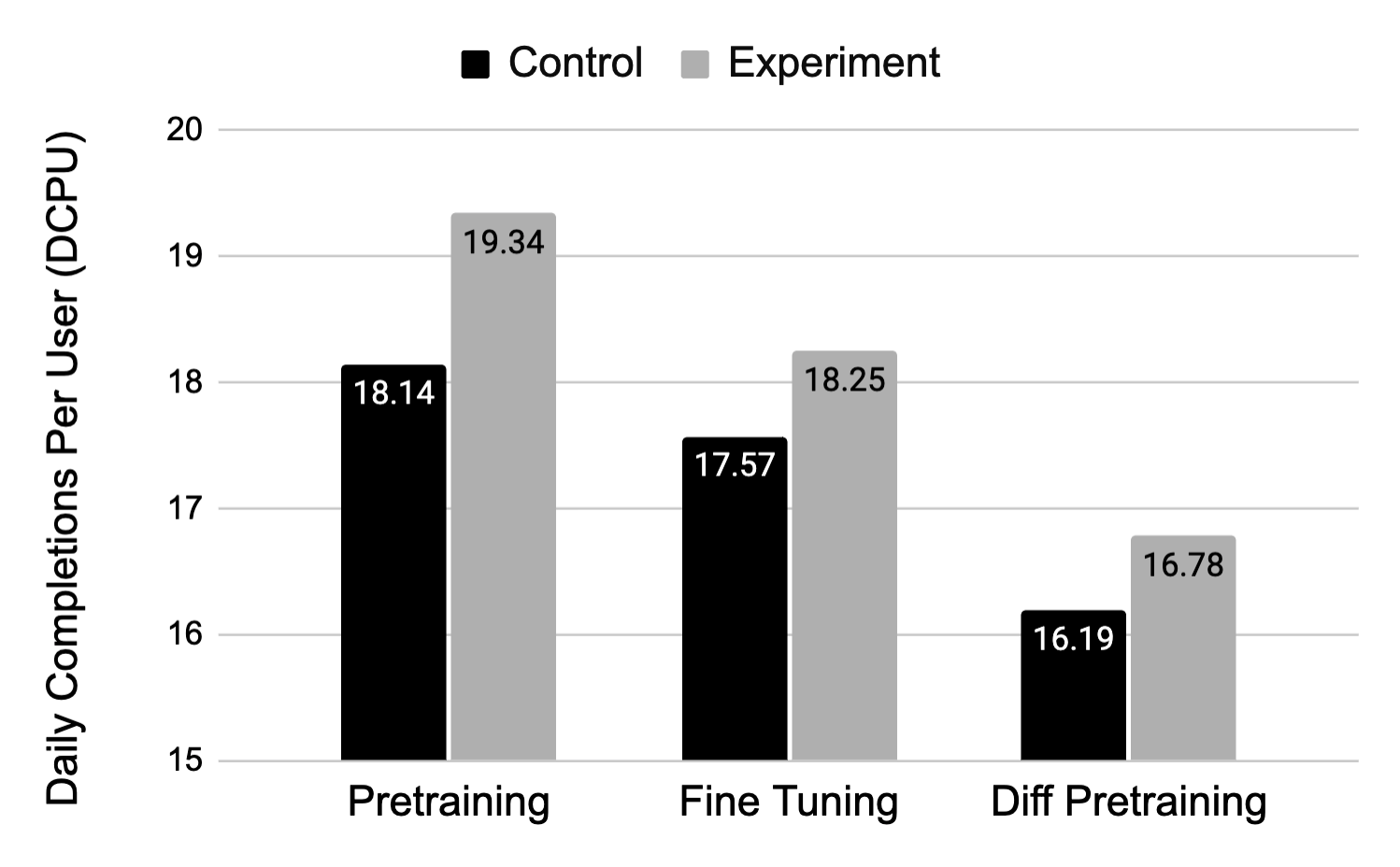}
\caption{Production A/B test results of three experiments. In the ``pretraining" test the experiment group model is pretrained on \hackctx{} and the control group model is not pretrained. In the ``fine-tuning" test the experiment group is fine-tuned on \hackselec{} and the control group is not fine-tuned. In the final ``diff pretraining" test the experiment group model is pretrained on version control code commits and the control group model is not pretrained. For all tests, the experiment group had greater DCPU at a statistically significant p-value.}
\label{fig:ab-graph}
\end{figure}

\begin{leftbar}
\noindent \textit{RQ1: How do autocompletion models benefit from combining unsupervised pretraining with task-specific fine-tuning? How does their performance improve across offline and online evaluation?}
\end{leftbar}

\textbf{\emph{Offline evaluation.}}
Autocompletion models fine-tuned on labeled data (\hackselec{}) outperform models without task-specific fine-tuning across offline and online evaluation. In offline evaluation, Table \ref{tab:hack-offline-gpt2} rows 2-3 show that fine-tuning led to a top-1 accuracy increase from 39.73\% to 41.91\% (5.5\% improvement) for the GPT-2 model. 

When labeled examples were mixed into pretraining (\hackall{}), top-1 accuracy jumped from 40.25\% to 41.6\% (3.4\% improvement) as shown in Table \ref{tab:hack-offline-gpt2} rows 4-5. For \snippetbart{}, top-1 accuracy jumped from 44.91\% to 53.23\% (18.5\% improvement) as shown in Table \ref{tab:configs-results} row 2 vs. 6. 

The same trends were observed when training Python autocompletion models (Table \ref{tab:python-offline-gpt2} rows 2-3 and 4-5). 

\textbf{\emph{Online evaluation.}}
For online evaluation, we trained a GPT-2 autocompletion model on \hackctx{}. The experiment variant was then fine-tuned on \hackselec{} whereas the control variant did not undergo fine-tuning (same setting as in offline evaluation Table \ref{tab:hack-offline-gpt2} rows 2-3). Experiment 1 in table \ref{tab:ab-results} (visualized in figure \ref{fig:ab-graph}) shows that daily completions per user (DCPU) was 3.86\% greater in the experiment group at \emph{p} = 0.0238, consistent with the improvement in offline evaluation.

We conducted a second A/B experiment comparing the better model from the previous experiment (pretraining on \hackctx{} and fine-tuning on \hackselec{}) to a model trained on \hackselec{} without pretraining (same setting as in offline evaluation Table \ref{tab:hack-offline-gpt2} rows 3 and 1). Experiment 2 in table \ref{tab:ab-results} (visualized in figure \ref{fig:ab-graph}) shows an even larger improvement over the model without pretraining\textemdash the experiment group DCPU was 6.63\% greater at \emph{p} = 0.017.

% ===========================================================================================

\begin{leftbar}
\noindent\textit{RQ2: What is the effect of pretraining on a large source code dataset obtained from outside of code authoring? Can pretrained software language models be fine-tuned on IDE autocompletion to achieve better accuracy with fewer real-world examples?}
\end{leftbar}

\textbf{\emph{Offline evaluation.}}
Given limited real-world examples of IDE autocompletion, pretraining on version control commits can make a big difference. However, as the number of autocompletion examples grows, the benefit of pretraining diminishes. These observations held constant across the GPT-2 and \snippetbart{} models we pretrained on a commits dataset.

%To answer this question, we pretrain models on commit data (from outside the IDE) before %fine-tuning on a variable number of IDE autocompletion examples. 

%While the code from autocomplete data gives us supervised learning training data of what %developers actually clicked on, diff data can be practically easier to obtain since %historically there are much more code changes data - this means diff data can provide %significantly more training data. Thus, to answer this question, we answered two questions - %if the number of training datapoints remained relatively similar, what are the implications %between the two models? And if fewer real-world examples were available, would having the %diff data be sufficient enough to achieve similar accuracy score?

Table \ref{tab:configs-results} row 5 vs. 3 shows that \diff{} pretraining with \hackselec{} fine-tuning outperforms \hackselec{} by 3.29\% top-1 accuracy (39.61\% vs. 36.32\%). The top-1 accuracy improvement is even greater for \snippetbart{}: 6.52\% (51.06\% vs. 44.54\%). However, pretraining on \diff{} yields worse performance compared to pretraining on IDE code sequences (\hackctx{}) as shown in row 5-6 in Table \ref{tab:configs-results}. The GPT-2 variant has 1.99\% lower top-1 accuracy (39.61\% vs. 41.60\%) whereas the \snippetbart{} variant is 2.17\% weaker (51.06\% vs. 53.23\%). This result aligned with our expectations as the \diff{} dataset, being sourced from a different stage of software development, exhibits greater concept drift from autocompletion than code sequences randomly sampled during code authoring.

\begin{figure}[h!]
\includegraphics[width=0.95\linewidth]{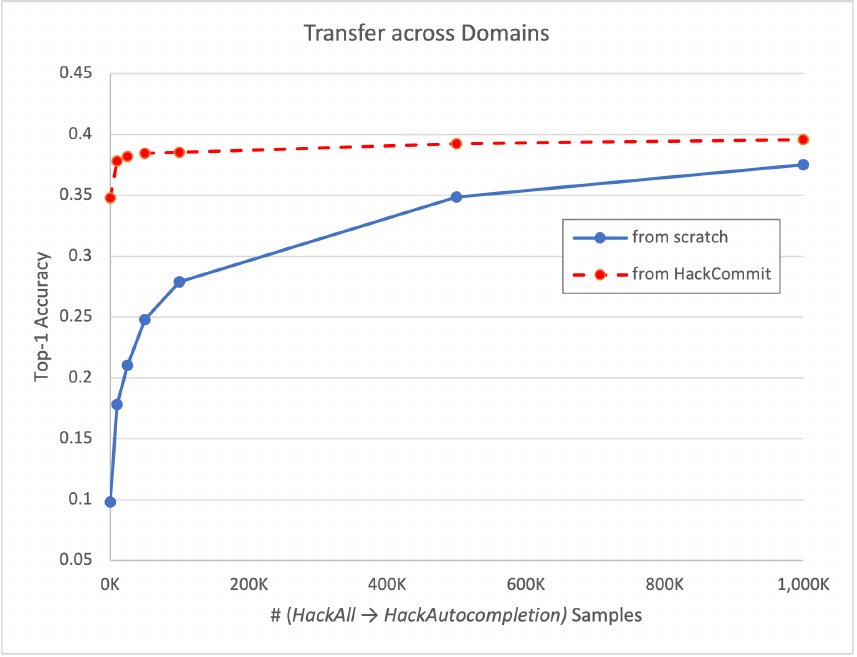}
\caption{Offline top-1 accuracy evaluation starting at \diff{} checkpoint vs random initialization as size of (\hackall{} $\rightarrow$ \hackselec{}) fine-tuning samples increases}
\label{fig: domain_transfer}
\end{figure}

We also experimented with stacking the two pretrainings (first \diff{} and then \hackall{}) before fine-tuning on \hackselec{} and found that these models, whether GPT-2 or \snippetbart{}, did not show meaningful improvement over a single \hackall{} pretraining. To understand why multiple pretrainings did not result in better performance, we conducted a final experiment in which the number of real-world fine-tuning examples was varied from 0\% to 100\%. The results are shown in Figure \ref{fig: domain_transfer}.

What we found is that pretraining on \diff{} has a diminishing marginal benefit as the number of fine-tuning real-world completion samples grows. Given a small number of real-world examples, pretraining on \diff{} has a major impact. For example, there is an improvement of 17.25\% (37.48\% vs. 20.23\%) when we limited the number of autocompletion examples to 25k! However, at a certain point, given enough training data points drawn from the same distribution as our evaluation dataset, pretraining on a dataset with domain concept drift is no longer helpful.

\textbf{\emph{Online evaluation.}}
We conducted an A/B experiment to explore the real-world impact of pretraining on code commits. The GPT-2 model trained on 100k IDE + Autocompletion samples was compared against a variant pretrained on all of the data from \diff{} and fine-tuned on 100k IDE + Autocompletion samples. Experiment 3 in table \ref{tab:ab-results} (visualized in figure \ref{fig:ab-graph}) shows that the pretrained model drove an improvement of 3.62\% DCPU at \emph{p} = 0.049.

% \textbf{\emph{Experiment - varying number of fine-tuned examples.}}
% We also experimented with stacking the two pretrainings (first \diff{} and then \hackall{}) before fine-tuning on \hackselec{} and found that these models, whether GPT-2 or \snippetbart{}, did not show meaningful improvement over a single \hackall{} pretraining. To understand why multiple pretrainings did not result in better performance, we conducted a final experiment in which the number of real-world fine-tuning examples was varied from 0\% to 100\%. The results are shown in Figure 3.

% What we found is that pretraining has a diminishing marginal benefit as the number of fine-tuning examples grows. Given a small number of real-world examples, pretraining on \diff{} has a major impact. For example, there is an improvement of 17.25\% (37.48\% vs. 20.23\%) when we limited the number of autocompletion examples to 25k! However, at a certain point, given enough training data points drawn from the same distribution as our evaluation dataset, pretraining on a dataset with domain concept drift is no longer helpful.

\begin{table*}[]
\centering
\def\arraystretch{1.20}
\caption{Offline evaluations of mono-lingual GPT-2 models on Hack}
\begin{tabular*}{0.8\textwidth}{l@{\extracolsep{\fill}}llrrrr}
\toprule
 & Pretraining & Fine-tuning & \multicolumn{1}{l}{Convergence Epochs} & \multicolumn{1}{l}{Top1 Acc} & \multicolumn{1}{l}{Top3 Acc} & \multicolumn{1}{l}{MRR} \\ \midrule
1 & - & \hackselec{}         & 6     & 0.3632          & 0.5614          & 0.4508          \\
2 & \hackctx{}   & -         & 18    & 0.3973          & 0.5816          & 0.4787          \\
3 & \hackctx{}   & \hackselec{} & 18+11 & \textbf{0.4191} & \textbf{0.5987} & \textbf{0.4988} \\
4 & \hackall{}   & -         & 12    & 0.4025          & 0.5858          & 0.4835          \\
5 &\hackall{}   & \hackselec{} & 12+7  & 0.4160          & 0.5970          & 0.4963          \\ \midrule

6 & \diff{}   & -             & 20 & 0.3479          & 0.5357        & 0.4306         \\
7 & \diff{}   & \hackselec{}  & 20+4 & 0.3961          & 0.5857        & 0.4801         \\
8 & \diff{}   & \hackall{}    & 20+20 & 0.4026          & 0.5867        & 0.4841         \\ 
9 & \diff{}   & \hackall{} $\rightarrow$ \hackselec{} & 20+20+20 & 0.4145        & 0.5962          & 0.4953          \\ \bottomrule
\end{tabular*}
\label{tab:hack-offline-gpt2}
\end{table*}

\begin{table*}[]
\centering
\def\arraystretch{1.20}
\caption{Offline evaluations of mono-lingual GPT-2 models on Python}
\begin{tabular*}{0.8\textwidth}{l@{\extracolsep{\fill}}llrrrr}
\toprule
& Pretraining & Fine-tuning & \multicolumn{1}{l}{Convergence Epochs} & \multicolumn{1}{l}{Top1 Acc} & \multicolumn{1}{l}{Top3 Acc} & \multicolumn{1}{l}{MRR} \\ \midrule
1 & - & \pythonselec{}          & 6    & 0.5282 & 0.6951 & 0.6029 \\
2 & \pythonctx{}   & -           & 15   & 0.5610 & 0.7228 & 0.6331 \\
3 & \pythonctx{}     & \pythonselec{} & 15+13                      & \textbf{0.5751}              & \textbf{0.7286}              & \textbf{0.6439}         \\
4 & \pythonall{}   & -           & 19   & 0.5605 & 0.7188 & 0.6313 \\
5 &\pythonall{}   & \pythonselec{} & 19+8 & 0.5723 & 0.7253 & 0.6408 \\ \bottomrule
\end{tabular*}
\label{tab:python-offline-gpt2}
\end{table*}

% ===========================================================================================

\begin{leftbar}
\noindent\textit{RQ3: Consider the case where a large training corpus is available in one language but not another. Can pretraining a multilingual model on the language with more training data benefit the language with less data?}
\end{leftbar}

\textbf{\emph{Offline evaluation.}}
We first combined the Hack and Python corpora to investigate whether a larger, diverse pretraining would improve performance. In addition to incorporating multiple languages in pretraining, we tested fine-tuning on examples from one language against fine-tuning on examples from multiple. Fine-tuning on a union of Hack and Python examples \multiselec{}{} led to the best-performing multilingual model across Hack and Python evaluation. Table \ref{fig:multi-results} shows improvements of 0.5\% and 1.34\% above fine-tuning on \hackselec{} and \pythonselec{} respectively. However, none of the multilingual models showed significant improvement over the best monolingual models. The best multilingual model had 0.53\% better top-1 accuracy than the best Python model but showed 0.02\% worse top-1 accuracy than the best Hack model. We hypothesize that combining programming language examples across languages has a diminishing marginal benefit as the number of examples available in each language grows.

To verify this hypothesis, we pretrain GPT-2 models on \hackall{}, fine-tune on varying amounts of \pythonall{}, and evaluate on held-out \pythonselec{} examples. The baselines we use for comparison are models with the same configuration trained on an equal number of \pythonall{} examples without any pretraining. This experiment was designed to show whether models pretrained on Hack data exhibited superior prediction performance on Python autocompletion examples, indicating knowledge transfer across programming languages.

\begin{figure}[h!]
\includegraphics[width=0.95\linewidth]{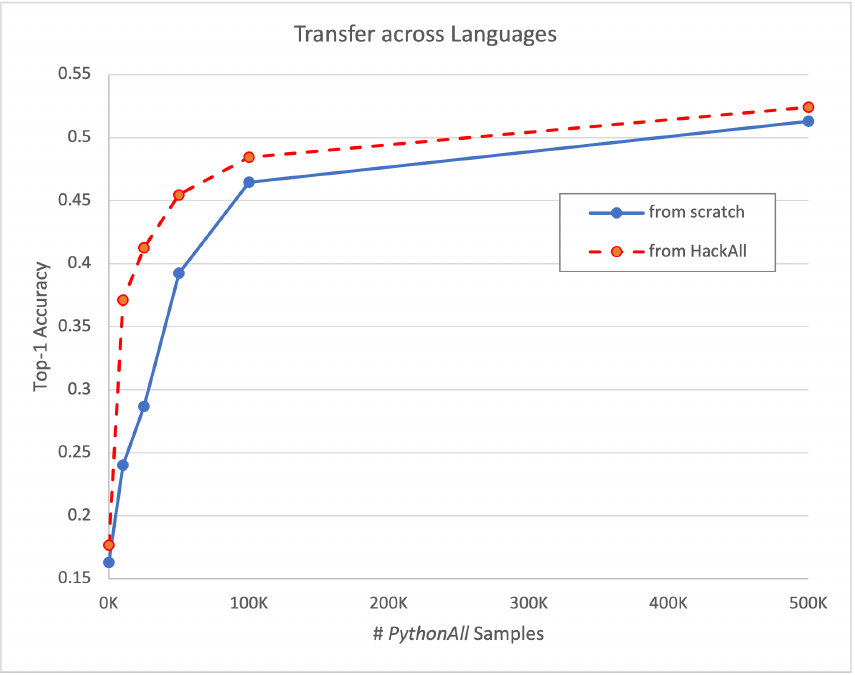}
\caption{Offline top-1 accuracy evaluation starting at \hackall{} checkpoint vs random initialization as size of \pythonall{} fine-tuning samples increases}
\label{fig: language_transfer}
\end{figure}

Figure \ref{fig: language_transfer} shows that the models pretrained on \hackall{} had better performance independent of the number of \pythonall{} examples used in fine-tuning. The marginal impact was greatest when we limited models to only 10k (13.1\% better top-1 accuracy, 37.11\% vs. 24.01\%) and 25k (12.6\% better top-1 accuracy, 41.26\% vs. 28.66\%) \pythonall{} examples. 
% Starting from a strong Hack solution (as opposed to random parameter initialization) helped GPT-2 find a good Python solution with drastically fewer examples. 
This shows clear evidence of knowledge transfer across programming languages in autocompletion. 
The performance of the model pretrained on \hackall{} and fine-tuned with 25k and 50k \pythonall{} examples is similar to the performance of training from scratch on 50k and 100k \pythonall{} examples, respectively. This shows that half as many examples were required for comparable performance after pretraining on an other-language dataset.

This is a meaningful insight for IDE autocompletion developers. Consider the case of providing predictive autocompletion ranking for less common programming languages (or ones for which real-world training examples are scarce). Our results show that pretraining on real-world examples from other languages makes it possible to achieve high accuracy with a relatively small fine-tuning dataset.

\textbf{\emph{Online evaluation.}} Our multilingual model's vocabulary was twice as large as either of the monolingual model vocabularies because it combines the vocabularies of two languages. Because language model latency is highly sensitive to vocabulary size, we could not perform a fair online A/B test comparing a multilingual source code model to a monolingual one under our experiment configuration. The prediction latency for the multilingual model was too high.

\begin{table*}[hbt]
\def\arraystretch{1.20}
\caption{Offline evaluations of multi-lingual GPT-2 models on Hack and Python}
\begin{tabular*}{\textwidth}{l@{\extracolsep{\fill}}lrrrrrrr}
\toprule
\multirow{2}{*}{Pretraining} &
  \multirow{2}{*}{Fine-tuning} &
  \multicolumn{1}{l}{\multirow{2}{*}{Epochs}} &
  \multicolumn{3}{c}{Hack} &
  \multicolumn{3}{c}{Python} \\ \cline{4-6} \cline{7-9} 
 &
   &
  \multicolumn{1}{l}{} &
  \multicolumn{1}{l}{Top1 Acc} &
  \multicolumn{1}{l}{Top3 Acc} &
  \multicolumn{1}{l}{MRR} &
  \multicolumn{1}{l}{Top1 Acc} &
  \multicolumn{1}{l}{Top3 Acc} &
  \multicolumn{1}{l}{MRR} \\ \midrule
- & \multiselec{}        & 7     & 0.3690 & 0.5677 & 0.4569 & 0.5416 & 0.7037 & 0.6142 \\
\hline
\multictx{}   & -           & 20    & 0.3981 & 0.5825 & 0.4796 & 0.5614 & 0.7260 & 0.6348 \\
\multictx{}  & \hackselec{}   & 20+14 & 0.4178 & 0.5973 & 0.4975 & 0.4951 & 0.6829 & 0.5780 \\
\multictx{}  & \pythonselec{} & 20+20 & 0.3616 & 0.5541 & 0.4465 & 0.5746 & 0.7274 & 0.6429 \\
\multictx{}  & \multiselec{}  & 20+12 & 0.4187 & 0.5987 & 0.4986 & 0.5759 & 0.7278 & 0.6439 \\
\hline
\multiall{}  & -           & 14    & 0.4046 & 0.5882 & 0.4859 & 0.5688 & 0.7302 & 0.6408 \\
\multiall{}  & \hackselec{}   & 14+10 & 0.4178 & 0.5977 & 0.4977 & 0.5321 & 0.7042 & 0.6086 \\
\multiall{}  & \pythonselec{} & 14+12 & 0.3703 & 0.5584 & 0.4532 & 0.5791 & 0.7297 & 0.6466 \\
\multiall{} & \multiselec{} &  14+9 &
  \textbf{0.4190} &
  \textbf{0.5988} &
  \textbf{0.4988} &
  \textbf{0.5804} &
  \textbf{0.7308} &
  \textbf{0.6478} \\ \bottomrule
\end{tabular*}
\label{fig:multi-results}
\end{table*}

\begin{table*}[]
\centering
\def\arraystretch{1.20}
\caption{Offline evaluations of GPT-2 and \snippetbart{} in each of 7 configs on Hack}
\begin{tabular*}{\textwidth}{l@{\extracolsep{\fill}}lrrrrrr}
% \toprule
%   & Config             & \multicolumn{1}{l}{Top1 Acc} & \multicolumn{1}{l}{Top3 Acc} & \multicolumn{1}{l}{MRR}  & \multicolumn{1}{l}{Top1 Acc} & \multicolumn{1}{l}{Top3 Acc} & \multicolumn{1}{l}{MRR} \\ \midrule
\toprule
\multirow{2}{*}{} &
  \multirow{2}{*}{Config} &
  \multicolumn{3}{c}{GPT-2} &
  \multicolumn{3}{c}{\snippetbart{}} \\ \cline{3-5} \cline{6-8} 
 & &
  \multicolumn{1}{l}{Top1 Acc} &
  \multicolumn{1}{l}{Top3 Acc} &
  \multicolumn{1}{l}{MRR} &
  \multicolumn{1}{l}{Top1 Acc} &
  \multicolumn{1}{l}{Top3 Acc} &
  \multicolumn{1}{l}{MRR} \\ \midrule

1 & \diff{}                       & 0.3479 & 0.5357 & 0.4306 & 0.4280 & 0.6608 & 0.5312 \\
2 & \hackall{}                 & 0.4025 & 0.5858 & 0.4835  & 0.4491 & 0.6863 & 0.5545 \\
3 & \hackselec{}               & 0.3632 & 0.5614 & 0.4508 & 0.4454 & 0.6806 & 0.5498 \\
4 & \diff{} $\rightarrow$ \hackall{}        & 0.4026 & 0.5867 & 0.4841  & 0.4471 & 0.6820 & 0.5514 \\
5 & \diff{} $\rightarrow$ \hackselec{}        & 0.3961 & 0.5857 & 0.4801  & 0.5106 & 0.7315 & 0.6096 \\
6 & \hackall{} $\rightarrow$ \hackselec{}  & \textbf{0.4160} & \textbf{0.5970} & \textbf{0.4963} & \textbf{0.5323} & \textbf{0.7497} & \textbf{0.6296} \\
7 & \diff{}$\rightarrow$\hackall{}$\rightarrow$\hackselec{}  & 0.4145  & 0.5962 & 0.4953  & \textbf{0.5323} & 0.7490 & 0.6293   \\ \bottomrule
\end{tabular*}
\label{tab:configs-results}
\end{table*}

% \begin{table*}[]
% \centering
% \def\arraystretch{1.10}
% \caption{Offline evaluations of \snippetbart{} in each of 7 configs on Hack}
% \begin{tabular}{@{}llrrr@{}}
% \toprule
%   & Config                              & \multicolumn{1}{l}{Top1 Acc} & \multicolumn{1}{l}{Top3 Acc} & \multicolumn{1}{l}{MRR} \\ \midrule
% 1 & \diff{}                       & 0.428 & 0.6608 & 0.5312 \\
% 2 & \hackall{}                & 0.4491 & 0.6863 & 0.5545 \\
% 3 & \hackselec{}                & 0.4454 & 0.6806 & 0.5498 \\
% 4 & \diff{} $\rightarrow$ \hackall{}        & 0.4471 & 0.682 & 0.5514 \\
% 5 & \diff{} $\rightarrow$ \hackselec{}        & 0.5106 & 0.7315 & 0.6096 \\
% 6 & \hackall{} $\rightarrow$ \hackselec{} & 0.5323 & 0.7497 & 0.6296 \\
% 7 & \diff{}$\rightarrow$\hackall{}$\rightarrow$\hackselec{} & 0.5323 & 0.749 & 0.6293                  \\ \bottomrule
% \end{tabular}
% \end{table*}

\section{Related Work}
\label{sec:relatedwork}

\noindent \textbf{IDE Autocompletion.}
Many studies at the intersection of software development tools and machine learning have investigated next code token prediction and its application to autocompletion.
Earlier attempts at modeling software languages and completion
ranking were based on n-gram language models~\cite{Hindle:2012:NS:2337223.2337322,nguyen2013statistical,hellendoorn2017modeling} or probabilistic models using the information from ASTs~\cite{bielik2016phog}.
With the advancement of deep learning models, RNNs (and their variants)~\cite{Li_2018,DBLP:journals/corr/abs-1903-05734} have shown promising improvements.
More recently, Transformers have achieved state-of-the-art performance for software language modeling~\cite{kim2020code,brockschmidt2018generative}.
Galois (Radford et al., 2019) and TabNine are two additional code completion tools that employ the GPT-2 Transformer for next token prediction.
Furthermore, some studies have focused on the industry use case of autocompletion for IDE users within a company.
While ~\cite{aye2020sequence} showed that a pointer mixture network performs well on an open-source GitHub corpus as well as Google's internal Dart language corpus, ~\cite{hellendoorn-codefails} warns that accuracy achieved on synthetic benchmarks may not translate to real-world completion performance.
Aye et al.~\cite{aye2020learning} demonstrated how training on real-world developer activity can combat this challenge.

\break

\noindent \textbf{Transfer Learning for Code.}
Transfer learning has revolutionized natural language processing (NLP) since the seminal papers on models such as GPT-2~\cite{radford2019language}, BERT~\cite{devlin2018bert} and RoBERTa~\cite{liu2019roberta}.
These works proposed the idea of pretraining Transformer models on large, diverse text corpora in an unsupervised manner and fine-tuning them for specific downstream tasks.
Since then several works have been proposed~\cite{feng2020codebert,mashhadi2021applying,lin2021traceability,SHARMA2021110936,mastropaolo2021studying,pei2021trex,elnaggar2021codetrans,ahmad2020summarization,lu2021codexglue} to apply the idea of transfer learning in the domain of code.

CodeBERT~\cite{feng2020codebert} was one of the first models pretrained on pairs of code and natural language sequences in order learn a bimodal representation of both entities.
It was applied on natural language code search and code documentation generation to demonstrate the benefit of transfer learning compared to training directly on the downstream tasks. A follow-up study~\cite{mashhadi2021applying} showed that CodeBERT can also transfer knowledge to the problem of automated program repair. More recently, \snippetbart{}~\cite{ahmad2020summarization} showed that the BART architecture~\cite{lewis-etal-2020-bart} is better-suited for source code generation tasks compared to BERT-based models.

In comparison to these works, the goal of our paper is not to produce a state-of-the-art model for any specific task. While there is at least one effort~\cite{lu2021codexglue} to create a multitask benchmark of source code modeling tasks in the style of the natural language decathlon\cite{mccann2018natural}, it is out of scope for us to compare the efficacy of our off-the-shelf models. Rather it is to show how transfer learning across various dimensions can benefit IDE autocompletion prediction {\em regardless of the model}.

In our experiments we used two state-of-the-art models \trans{} and \snippetbart{} for this purpose, and showed that both models can benefit from transfer learning.

In a closely related work, Mastropaolo et al.~\cite{mastropaolo2021studying} study empirically how the T5 model~\cite{2020t5} behaves when pretrained and fine-tuned on four code-related tasks. They also make an observation that an unsupervised pretraining phase helps the model achieve better performance on their set of tasks, such as code summarization and mutation.
In our paper, in addition to observing this effect in the downstream task of IDE autocompletion prediction, we also explored transfer learning in other dimensions, such as across programming languages, and validated our results through online A/B tests.

Outside of code authoring, transfer learning has been enabled by Transformer-based models in other software artifacts. Lin et al.~\cite{lin2021traceability} apply pretrained BERT models for learning relationships between issues and commits in a software repository. Sharma et al.~\cite{SHARMA2021110936} detect code smells in programming languages where sufficient training data is not available by transferring knowledge from other data-rich languages. Pei et al.~\cite{pei2021trex} propose a tool called TREX that is better at detecting binary-level function similarity than state-of-the-art tools owing to transfer learning from binary-level execution traces.

\section{Threats to Validity}
\label{sec:threats}

\paragraph{Intersection of vocabulary between languages} In our investigation of multilingual software language modeling, one method we used to reduce the out-of-vocabulary (OOV) problem was to tokenize code sequences using a bigram encoding. Although this reduced the OOV rate for individual languages, there was only a small overlap between the different languages' vocabularies. Applying BPE or another encoding scheme may have resulted in more tokens receiving the same encoding across languages which could increase the efficacy of transfer learning across programming languages.

\paragraph{Testing on more languages} We examined the effects of transfer learning on two languages: Hack and Python. While we saw that transfer learning (over various dimensions) improves performance, we did not evaluate on other languages. Hack and Python are both dynamic, object-oriented programming languages. It's conceivable that knowledge transfer to a static language like C++ or a different paradigm language like OCaml could be less effective.

\paragraph{Facebook vs open-source} When exploring pretraining on code commits from version control, we leveraged Facebook's internal version control repository. It's possible that some of the transfer learning effects we observed when fine-tuning on IDE autocompletion would be lessened if we had instead pretrained on GitHub code commits. In addition to the greater source code diversity in the GitHub corpus, there is undoubtedly concept drift between code in open-source and at Facebook.  There may even be different version control usage patterns that would affect pretraining on code commits.

\section{Future Work}
\label{sec:future}

\paragraph{Stronger ablation study on PLBART} For both languages, PLBART outperformed the GPT-2 model by more than 10\%. However, it is difficult to make an apples-to-apples comparison. A bidirectional model, PLBART leverages context after the predicted token while GPT-2 only uses the before-cursor context. PLBART also uses BPE instead of a bigram + copy mechanism encoding. In the future we wish to do a more thorough ablation study to determine the biggest contributing factors to PLBART's performance.

\paragraph{Transfer learning across multiple languages} In this paper, we focused on transfer learning between two languages: Hack and Python. However, there are many other languages that are used in software engineering. Our experiments showed that pretraining on one language transfers knowledge to the other. Does the impact of transfer learning grow or diminish as we add more languages to pretraining? Could pretraining on multiple languages decrease even further the number of fine-tuning examples needed in the target language?

\paragraph{Transfer learning across source code tasks} The evaluation task for this paper was IDE autocompletion prediction. Could we leverage our pretrained models for transfer learning to other source code tasks such as as bug fixing or similar code detection? Furthermore, could the pretrained model for another task be used effectively as the base model for autocompletion as well?
\section{Conclusion}
\label{sec:conclusion}

In this paper, we explored ways in which transfer learning can improve autocompletion. Previous work showed that a training corpus consisting of real-world examples collected from programmers' IDE usage leads to the highest accuracy autocompletion models. But for some tool developers, there may be a limited number of real-world autocompletion examples in the desired programming language. This study showed how the power of transfer learning enables pretraining on non-IDE, non-autoompletion, and different-language example code sequences before fine-tuning on the autocompletion prediction task. Our results show that we can reach comparable accuracy while drastically reduce the number of fine-tuning examples by starting from a pretrained model. These findings in offline evaluation were confirmed in online A/B experiments conducted on thousands of software developers at Facebook. 

% \newpage
\bibliographystyle{IEEEtran}
\bibliography{references.bib}

\begin{thebibliography}{10}
\providecommand{\url}[1]{#1}
\csname url@samestyle\endcsname
\providecommand{\newblock}{\relax}
\providecommand{\bibinfo}[2]{#2}
\providecommand{\BIBentrySTDinterwordspacing}{\spaceskip=0pt\relax}
\providecommand{\BIBentryALTinterwordstretchfactor}{4}
\providecommand{\BIBentryALTinterwordspacing}{\spaceskip=\fontdimen2\font plus
\BIBentryALTinterwordstretchfactor\fontdimen3\font minus
  \fontdimen4\font\relax}
\providecommand{\BIBforeignlanguage}[2]{{%
\expandafter\ifx\csname l@#1\endcsname\relax
\typeout{** WARNING: IEEEtran.bst: No hyphenation pattern has been}%
\typeout{** loaded for the language `#1'. Using the pattern for}%
\typeout{** the default language instead.}%
\else
\language=\csname l@#1\endcsname
\fi
#2}}
\providecommand{\BIBdecl}{\relax}
\BIBdecl

\bibitem{aye2020learning}
G.~A. Aye, S.~Kim, and H.~Li, ``Learning autocompletion from real-world
  datasets,'' 2020.

\bibitem{Murphy:2006:JSD:1159169.1159396}
\BIBentryALTinterwordspacing
G.~C. Murphy, M.~Kersten, and L.~Findlater, ``How are java software developers
  using the eclipse ide?'' \emph{IEEE Softw.}, vol.~23, no.~4, pp. 76--83, Jul.
  2006. [Online]. Available: \url{http://dx.doi.org/10.1109/MS.2006.105}
\BIBentrySTDinterwordspacing

\bibitem{Bruch:2009:LEI:1595696.1595728}
\BIBentryALTinterwordspacing
M.~Bruch, M.~Monperrus, and M.~Mezini, ``Learning from examples to improve code
  completion systems,'' in \emph{Proceedings of the the 7th Joint Meeting of
  the European Software Engineering Conference and the ACM SIGSOFT Symposium on
  The Foundations of Software Engineering}, ser. ESEC/FSE '09.\hskip 1em plus
  0.5em minus 0.4em\relax New York, NY, USA: ACM, 2009, pp. 213--222. [Online].
  Available: \url{http://doi.acm.org/10.1145/1595696.1595728}
\BIBentrySTDinterwordspacing

\bibitem{Hindle:2012:NS:2337223.2337322}
\BIBentryALTinterwordspacing
A.~Hindle, E.~T. Barr, Z.~Su, M.~Gabel, and P.~Devanbu, ``On the naturalness of
  software,'' in \emph{Proceedings of the 34th International Conference on
  Software Engineering}, ser. ICSE '12.\hskip 1em plus 0.5em minus 0.4em\relax
  Piscataway, NJ, USA: IEEE Press, 2012, pp. 837--847. [Online]. Available:
  \url{http://dl.acm.org/citation.cfm?id=2337223.2337322}
\BIBentrySTDinterwordspacing

\bibitem{aye2020sequence}
G.~A. Aye and G.~E. Kaiser, ``Sequence model design for code completion in the
  modern ide,'' 2020.

\bibitem{kim2020code}
S.~Kim, J.~Zhao, Y.~Tian, and S.~Chandra, ``Code prediction by feeding trees to
  transformers,'' 2020.

\bibitem{Li_2018}
\BIBentryALTinterwordspacing
J.~Li, Y.~Wang, M.~R. Lyu, and I.~King, ``Code completion with neural attention
  and pointer networks,'' \emph{Proceedings of the Twenty-Seventh International
  Joint Conference on Artificial Intelligence}, Jul 2018. [Online]. Available:
  \url{http://dx.doi.org/10.24963/ijcai.2018/578}
\BIBentrySTDinterwordspacing

\bibitem{hellendoorn-codefails}
\BIBentryALTinterwordspacing
V.~J. Hellendoorn, S.~Proksch, H.~C. Gall, and A.~Bacchelli, ``When code
  completion fails: A case study on real-world completions,'' ser. ICSE
  '19.\hskip 1em plus 0.5em minus 0.4em\relax IEEE Press, 2019, p. 960–970.
  [Online]. Available: \url{https://doi.org/10.1109/ICSE.2019.00101}
\BIBentrySTDinterwordspacing

\bibitem{radford2019language}
A.~Radford, J.~Wu, R.~Child, D.~Luan, D.~Amodei, and I.~Sutskever, ``Language
  models are unsupervised multitask learners,'' 2019.

\bibitem{devlin2018bert}
J.~Devlin, M.-W. Chang, K.~Lee, and K.~Toutanova, ``Bert: Pre-training of deep
  bidirectional transformers for language understanding,'' \emph{arXiv preprint
  arXiv:1810.04805}, 2018.

\bibitem{lewis-etal-2020-bart}
\BIBentryALTinterwordspacing
M.~Lewis, Y.~Liu, N.~Goyal, M.~Ghazvininejad, A.~Mohamed, O.~Levy, V.~Stoyanov,
  and L.~Zettlemoyer, ``{BART}: Denoising sequence-to-sequence pre-training for
  natural language generation, translation, and comprehension,'' in
  \emph{Proceedings of the 58th Annual Meeting of the Association for
  Computational Linguistics}.\hskip 1em plus 0.5em minus 0.4em\relax Online:
  Association for Computational Linguistics, Jul. 2020, pp. 7871--7880.
  [Online]. Available: \url{https://www.aclweb.org/anthology/2020.acl-main.703}
\BIBentrySTDinterwordspacing

\bibitem{feng2020codebert}
Z.~Feng, D.~Guo, D.~Tang, N.~Duan, X.~Feng, M.~Gong, L.~Shou, B.~Qin, T.~Liu,
  D.~Jiang, and M.~Zhou, ``Codebert: A pre-trained model for programming and
  natural languages,'' 2020.

\bibitem{ahmad2020summarization}
W.~U. Ahmad, S.~Chakraborty, B.~Ray, and K.-W. Chang, ``Unified pre-training
  for program understanding and generation,'' in \emph{Proceedings of the 2021
  Conference of the North {A}merican Chapter of the Association for
  Computational Linguistics}, 2021.

\bibitem{allamanis2019adverse}
M.~Allamanis, ``The adverse effects of code duplication in machine learning
  models of code,'' 2019.

\bibitem{keskar2019ctrl}
N.~S. Keskar, B.~McCann, L.~R. Varshney, C.~Xiong, and R.~Socher, ``Ctrl: A
  conditional transformer language model for controllable generation,'' 2019.

\bibitem{DBLP:journals/corr/abs-1903-05734}
\BIBentryALTinterwordspacing
R.~Karampatsis and C.~Sutton, ``Maybe deep neural networks are the best choice
  for modeling source code,'' \emph{CoRR}, vol. abs/1903.05734, 2019. [Online].
  Available: \url{http://arxiv.org/abs/1903.05734}
\BIBentrySTDinterwordspacing

\bibitem{DBLP:journals/corr/SennrichHB15}
\BIBentryALTinterwordspacing
R.~Sennrich, B.~Haddow, and A.~Birch, ``Neural machine translation of rare
  words with subword units,'' \emph{CoRR}, vol. abs/1508.07909, 2015. [Online].
  Available: \url{http://arxiv.org/abs/1508.07909}
\BIBentrySTDinterwordspacing

\bibitem{DBLP:journals/corr/abs-2010-12663}
\BIBentryALTinterwordspacing
N.~Chirkova and S.~Troshin, ``A simple approach for handling out-of-vocabulary
  identifiers in deep learning for source code,'' \emph{CoRR}, vol.
  abs/2010.12663, 2020. [Online]. Available:
  \url{https://arxiv.org/abs/2010.12663}
\BIBentrySTDinterwordspacing

\bibitem{nguyen2013statistical}
\BIBentryALTinterwordspacing
T.~T. Nguyen, A.~T. Nguyen, H.~A. Nguyen, and T.~N. Nguyen, ``A statistical
  semantic language model for source code,'' in \emph{Proceedings of the 2013
  9th Joint Meeting on Foundations of Software Engineering}, ser. ESEC/FSE
  2013.\hskip 1em plus 0.5em minus 0.4em\relax New York, NY, USA: Association
  for Computing Machinery, 2013, p. 532–542. [Online]. Available:
  \url{https://doi.org/10.1145/2491411.2491458}
\BIBentrySTDinterwordspacing

\bibitem{hellendoorn2017modeling}
\BIBentryALTinterwordspacing
V.~J. Hellendoorn and P.~Devanbu, ``Are deep neural networks the best choice
  for modeling source code?'' in \emph{Proceedings of the 2017 11th Joint
  Meeting on Foundations of Software Engineering}, ser. ESEC/FSE 2017.\hskip
  1em plus 0.5em minus 0.4em\relax New York, NY, USA: Association for Computing
  Machinery, 2017, p. 763–773. [Online]. Available:
  \url{https://doi.org/10.1145/3106237.3106290}
\BIBentrySTDinterwordspacing

\bibitem{bielik2016phog}
P.~Bielik, V.~Raychev, and M.~Vechev, ``Phog: Probabilistic model for code,''
  in \emph{Proceedings of the 33rd International Conference on International
  Conference on Machine Learning - Volume 48}, ser. ICML'16.\hskip 1em plus
  0.5em minus 0.4em\relax JMLR.org, 2016, p. 2933–2942.

\bibitem{brockschmidt2018generative}
M.~Brockschmidt, M.~Allamanis, A.~L. Gaunt, and O.~Polozov, ``Generative code
  modeling with graphs,'' \emph{arXiv preprint arXiv:1805.08490}, 2018.

\bibitem{liu2019roberta}
Y.~Liu, M.~Ott, N.~Goyal, J.~Du, M.~Joshi, D.~Chen, O.~Levy, M.~Lewis,
  L.~Zettlemoyer, and V.~Stoyanov, ``Roberta: A robustly optimized bert
  pretraining approach,'' 2019.

\bibitem{mashhadi2021applying}
E.~Mashhadi and H.~Hemmati, ``Applying codebert for automated program repair of
  java simple bugs,'' 2021.

\bibitem{lin2021traceability}
J.~Lin, Y.~Liu, Q.~Zeng, M.~Jiang, and J.~Cleland-Huang, ``Traceability
  transformed: Generating more accurate links with pre-trained bert models,''
  2021.

\bibitem{SHARMA2021110936}
\BIBentryALTinterwordspacing
T.~Sharma, V.~Efstathiou, P.~Louridas, and D.~Spinellis, ``Code smell detection
  by deep direct-learning and transfer-learning,'' \emph{Journal of Systems and
  Software}, vol. 176, p. 110936, 2021. [Online]. Available:
  \url{https://www.sciencedirect.com/science/article/pii/S0164121221000339}
\BIBentrySTDinterwordspacing

\bibitem{mastropaolo2021studying}
A.~Mastropaolo, S.~Scalabrino, N.~Cooper, D.~N. Palacio, D.~Poshyvanyk,
  R.~Oliveto, and G.~Bavota, ``Studying the usage of text-to-text transfer
  transformer to support code-related tasks,'' 2021.

\bibitem{pei2021trex}
K.~Pei, Z.~Xuan, J.~Yang, S.~Jana, and B.~Ray, ``Trex: Learning execution
  semantics from micro-traces for binary similarity,'' 2021.

\bibitem{elnaggar2021codetrans}
A.~Elnaggar, W.~Ding, L.~Jones, T.~Gibbs, T.~Feher, C.~Angerer, S.~Severini,
  F.~Matthes, and B.~Rost, ``Codetrans: Towards cracking the language of
  silicone's code through self-supervised deep learning and high performance
  computing,'' 2021.

\bibitem{lu2021codexglue}
S.~Lu, D.~Guo, S.~Ren, J.~Huang, A.~Svyatkovskiy, A.~Blanco, C.~Clement,
  D.~Drain, D.~Jiang, D.~Tang, G.~Li, L.~Zhou, L.~Shou, L.~Zhou, M.~Tufano,
  M.~Gong, M.~Zhou, N.~Duan, N.~Sundaresan, S.~K. Deng, S.~Fu, and S.~Liu,
  ``Codexglue: A machine learning benchmark dataset for code understanding and
  generation,'' 2021.

\bibitem{mccann2018natural}
B.~McCann, N.~S. Keskar, C.~Xiong, and R.~Socher, ``The natural language
  decathlon: Multitask learning as question answering,'' 2018.

\bibitem{2020t5}
\BIBentryALTinterwordspacing
C.~Raffel, N.~Shazeer, A.~Roberts, K.~Lee, S.~Narang, M.~Matena, Y.~Zhou,
  W.~Li, and P.~J. Liu, ``Exploring the limits of transfer learning with a
  unified text-to-text transformer,'' \emph{Journal of Machine Learning
  Research}, vol.~21, no. 140, pp. 1--67, 2020. [Online]. Available:
  \url{http://jmlr.org/papers/v21/20-074.html}
\BIBentrySTDinterwordspacing

\end{thebibliography}

\end{document}